\documentclass[%
 pra,
 amsmath,amssymb,
 reprint,%
]{revtex4-2}

\usepackage{amsmath,amsfonts}    
\usepackage{graphicx}   
\usepackage{verbatim}   
\usepackage{color,xcolor}      
\usepackage{subfigure}  
\usepackage{hyperref}   
\usepackage{textcomp}
\usepackage{bm}
\usepackage{bbm}
\usepackage{braket}
\usepackage[normalem]{ulem}

\graphicspath{{figs/}}

\draft 

\begin{document}


\title{Modifying quantum optical states by zero-photon subtraction} 



\author{C.M. Nunn}
\author{J.D. Franson}
\author{T.B. Pittman}
\affiliation{Department of Physics, University of Maryland Baltimore County, Baltimore, MD 21250, USA}


\date{\today}

\begin{abstract}
The process of single-photon subtraction (SPS) is known to dramatically alter the properties of certain quantum optical states. Somewhat surprisingly, subtracting zero photons can also modify quantum states and has practical applications in quantum communication. Here we experimentally investigate zero-photon subtraction (ZPS) using a wide variety of input states and conditional measurements based on actively detecting zero photons in one output port of a variable beamsplitter. We find that SPS and ZPS can exhibit complementary behavior depending on the photon statistics of the input states, and highlight deeper connections with Mandel's $Q$-parameter for classifying quantum states.
\end{abstract}


\maketitle 



\section{Introduction}\label{sec:intro}

In experimental quantum optics, the bosonic annihilation operator $\hat{a}$ can be realized through the process of ``single-photon subtraction'' (SPS)~\cite{wen_04,our_06}. When the input state contains a definite number of photons $\ket{n}$, this operation transforms the state as $\hat{a}\ket{n}\rightarrow\sqrt{n}\ket{n-1}$ in the usual way, corresponding to the simple removal of one photon from the state. However, when the input is a superposition of different number states, the SPS process can lead to counterintuitive results~\cite{ueda_90,miz_02}. For example, consider the input state $\ket{\psi}_{in}=\frac{1}{\sqrt{2}}\left(\ket{1}+\ket{5}\right)$, which has a mean number of photons $\langle \hat{n} \rangle = 3$ (where the number operator $\hat{n} \equiv \hat{a}^{\dagger} \hat{a}$). Applying $\hat{a}$ to this state leads to $\ket{\psi}_{out} = \frac{1}{\sqrt{6}}\left(\ket{0}+\sqrt{5}\ket{4}\right)$, which has $ \langle \hat{n} \rangle= 3.\bar{3}$. In this sense, subtracting a single photon from the state has actually {\em increased} the mean number of photons~\cite{par_07,zav_08}.
	
\begin{figure}[t]
	\centering
	\includegraphics[scale=0.50,trim={300 160 300 140},clip]{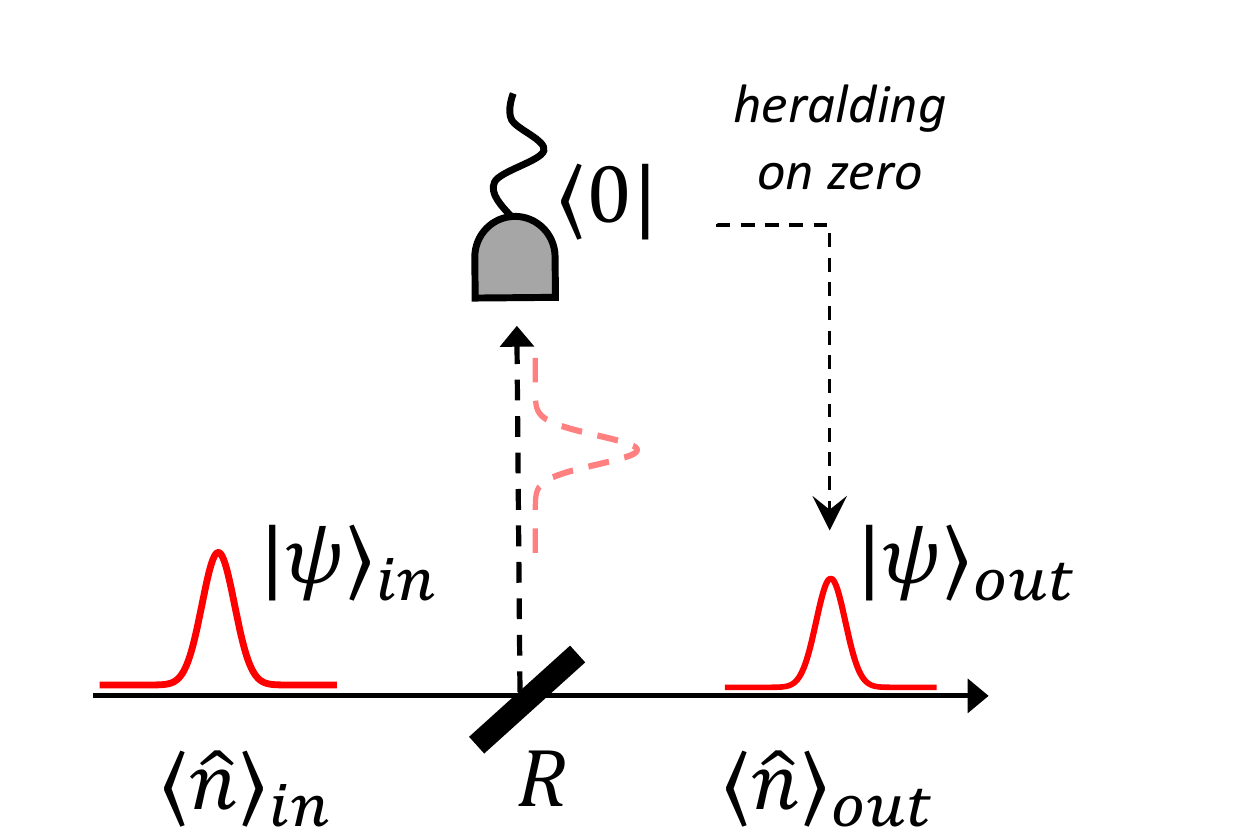}
	\caption{An implementation of ``zero-photon subtraction'' (ZPS) via conditional measurements on a beamsplitter. A superposition state $\ket{\psi}_{in}$ with expected photon number $\langle \hat{n}\rangle_{in}$ is prepared in the input mode of a beamsplitter with reflectance $R$. In contrast to single-photon subtraction (SPS), ZPS requires heralding on the detection of \textit{zero} photons in the reflected mode~\cite{nun_21}. Heralding on zero photons yields the attenuated state $\ket{\psi}_{out}$ with reduced mean photon number $\langle \hat{n} \rangle_{out}<\langle \hat{n}\rangle_{in}$. The degree of attenuation depends on both $R$ and the photon number statistics of the input state.}
	\label{fig:bsinout}
\end{figure}

In a similarly counterintuitive way, subtracting zero photons from a state can actually \textit{decrease} the mean number of photons. Figure~\ref{fig:bsinout} shows an implementation of this ``zero-photon subtraction'' (ZPS) process using a beamsplitter with reflectance $R$, and conditional measurements. A pulsed input state $\ket{\psi}_{in}$ passes through the beamsplitter, and the transmitted output $\ket{\psi}_{out}$ is heralded by the successful detection of zero photons in the reflected mode. Despite no photons being physically removed from the system, ZPS results in  $\langle \hat{n} \rangle_{out}<\langle \hat{n} \rangle_{in}$ for all but pure Fock states~\cite{gag_14}. Importantly, the ZPS process in Fig.~\ref{fig:bsinout} can be used to implement a probabilistic noiseless attenuation protocol that is useful for quantum communications~\cite{mic_12,ricky_17,guo_20}.

As highlighted by the structure of Fig.~\ref{fig:bsinout}, the key difference between SPS and ZPS is a heralding signal based on the detection of one vs. zero photons, respectively. While SPS has been experimentally studied extensively~\cite{bellini_10}, ZPS has only been briefly observed for super-Poissonian (i.e., thermal) states, and with fixed values of beamsplitter reflectance $R$~\cite{allevi_10,zhai_13,vid_16,bog_17,hlou_17,magl_19,kat_20}. In this paper, we systematically study ZPS for examples of super-Poissonian, sub-Poissonian, and coherent state inputs, all as a function of beamsplitter reflectance ranging from $R=0\rightarrow1$. The observed trends in attenuation demonstrate some complementary aspects of ZPS and SPS that depend on the photon number distributions, and highlight the role of losses and detector efficiency when heralding on zero photons in ZPS.

The remainder of the paper is structured as follows: in Section~\ref{sec:zps} we provide a detailed theoretical background for ZPS, and introduce an experimentally accessible parameter $K$ that can be used to quantify the degree of attenuation. In Section~\ref{sec:exp} we describe our experimental system, which uses (1) a conventional pulsed parametric down-conversion (PDC) source to produce the desired input states~\cite{gk_04}, (2) a variable evanescent-mode fiber coupler to continuously vary $R$~\cite{digshaw_82}, and (3) the ability to actively herald on zero photons using commercial single-photon detectors~\cite{nun_21}. In Section~\ref{sec:res} we analyze and discuss the experimental results, and briefly describe classical analogues that provide some additional insight into the observed attenuation effects. Finally, we summarize our study and conclude in Section~\ref{sec:con}.

\section{Zero-Photon Subtraction}\label{sec:zps}
The process of zero-photon subtraction (ZPS) illustrated in Figure~\ref{fig:bsinout} was first proposed as a method of noiseless attenuation by Mi\v{c}uda \textit{et al}.~\cite{mic_12}. This transformation can be defined by its action on Fock states $\ket{n}\rightarrow t^n\ket{n}$ with beamsplitter transmittance $T=|t|^2$. For an arbitrary input state $\hat\rho=\sum_{m,n=0}^\infty\rho_{mn}\ket{m}\bra{n}$, noiseless attenuation yields the following expected photon number in the output~\cite{gag_14}:
\begin{equation}\label{eqn:nout}
	\langle \hat{n} \rangle_{out} = \frac{\sum_n n \rho_{nn}T^n}{\sum_n \rho_{nn}T^n}
\end{equation}
When $T=1$, we regain the expected photon number of the original state with no attenuation, $\langle \hat{n} \rangle_{in}=\sum_n n \rho_{nn}$. Remarkably, when $T<1$ it can be seen that $\langle \hat{n} \rangle_{out}<\langle \hat{n} \rangle_{in}$ for all but pure Fock states by differentiating Eq.~\ref{eqn:nout} with respect to $T$:
\begin{equation}\label{eqn:dndt}
	\frac{d\langle \hat{n} \rangle_{out}}{dT} = \frac{1}{T}\left[\frac{\sum_n (n - \langle \hat{n} \rangle_{out})^2 \rho_{nn}T^n}{\sum_n \rho_{nn}T^n}\right] \equiv \frac{\langle (\Delta n)^2\rangle_{out}}{T} \geq 0,
\end{equation}
and seeing that $\langle \hat{n} \rangle_{out}$ increases monotonically on the interval $T\in(0,1]$. Here $\langle (\Delta n)^2\rangle_{out}$ is the photon number variance of the transformed state.

Equation~\ref{eqn:dndt} is analogous to the result derived by Ueda \textit{et al}. for stationary fields~\cite{ueda_90}, and it suggests that the degree of attenuation is closely related to the photon number statistics of the input state. Experimentally, it is convenient to quantify the degree of attenuation as a function of reflectance $R$ with the following ratio:
\begin{equation}\label{eqn:kdeff}
	K(R)\equiv \frac{\langle \hat{n} \rangle_{out}}{(1-R)\langle \hat{n} \rangle_{in}}
\end{equation}
The denominator $(1-R)\langle \hat{n} \rangle_{in}$ simply corresponds to ordinary attenuation by a beamsplitter, in which a fraction $T=1-R$ of the photons are transmitted on average. Thus, $K(R)$ compares the mean photon number of the heralded ZPS state $\langle \hat{n} \rangle_{out}$ to that of the ``ordinary'' output state with no conditional measurements.

The relative attenuation function $K(R)$ contains information about the photon number distribution and higher-order correlations. Most importantly, one can derive from Eqns.~\ref{eqn:nout} to~\ref{eqn:kdeff} that:
\begin{equation}\label{eqn:qin}
	\frac{dK}{dR}\Bigr|_{R=0}=1-\frac{\langle (\Delta n)^2\rangle_{in}}{\langle \hat{n} \rangle_{in}}\equiv -Q_{in}
\end{equation}
where $Q_{in}$ is Mandel's $Q$-parameter for the input state~\cite{man_79}.



The above result highlights an important connection between ZPS and typical SPS. In the limit of low beamsplitter reflectance $R$, SPS is equivalent to the annihilation operator $\hat{a}$~\footnote{Historically, ``photon subtraction'' refers to application of the annihilation operator, predating its realization by conditional measurements in the output of a beamsplitter (BS)~\cite{bellini_10}. However, the term often encompasses the more general BS transformations which relax the weak reflectance requirement (e.g.,~\cite{oli_03,kim_05}), which we follow here in our discussion of ZPS.}, and increases the mean photon number of some states as demonstrated in Sec.~\ref{sec:intro}. More precisely, this so-called ``photon excess'' is given exactly by the $Q$-parameter, such that $Q_{in}=\langle \hat{n}\rangle_{out}-\langle \hat{n} \rangle_{in}$~\cite{miz_02}. Thus, the mean photon number of super-Poissonian states $(Q>0)$ counterintuitively increases after performing SPS with a weakly reflecting beamsplitter. Equation~\ref{eqn:qin} links this property of SPS to the behavior of ZPS in the same regime of $R\ll 1$. For ZPS, the $Q$-parameter determines the initial slope $dK/dR$, and thus deviations from $K=1$ as $R$ increases from zero. We can therefore say super-Poissonian states exhibit a complementary ``photon deficit'' $(K<1)$ after ZPS in this regime, such that the mean photon number is reduced below that of ordinary attenuation. 

In the same way that SPS has unique consequences for sub-, super- and Poissonian states~\cite{zav_08}, it is also natural to investigate these three classes of states for ZPS. Our experiment will examine the following cases: (1) coherent states $\ket{\alpha}$, which possess Poissonian statistics; (2) the single-mode squeezed vacuum state (SMSV) $\ket{\xi}$, which is super-Poissonian; and (3) a single-photon Fock state $\ket{1}$, which is sub-Poissonian. As detailed in Section~\ref{sec:exp}, the experimentally prepared single-photon state is actually a mixture that includes the vacuum term, $\hat{\rho}_1=(1-\beta)\ket{0}\bra{0}+\beta\ket{1}\bra{1}$. We can calculate the expected relative attenuation for each of the three input states:
\begin{align}
	K^{(\alpha)}(R)&=1 \label{eqn:ka} \\
	K^{(\xi)}(R)&\approx 1-R \label{eqn:kx} \\
	K^{(\hat{\rho}_1)}(R)&=\frac{1}{1-\beta R} \label{eqn:kh} 
\end{align}
where the approximation for the SMSV $\ket{\xi}$ in Eq.~\ref{eqn:kx} holds for weak squeezing.

\section{Experiment}\label{sec:exp}

\begin{figure*}[t]
	\includegraphics[scale=0.55,trim={10 120 10 180},clip]{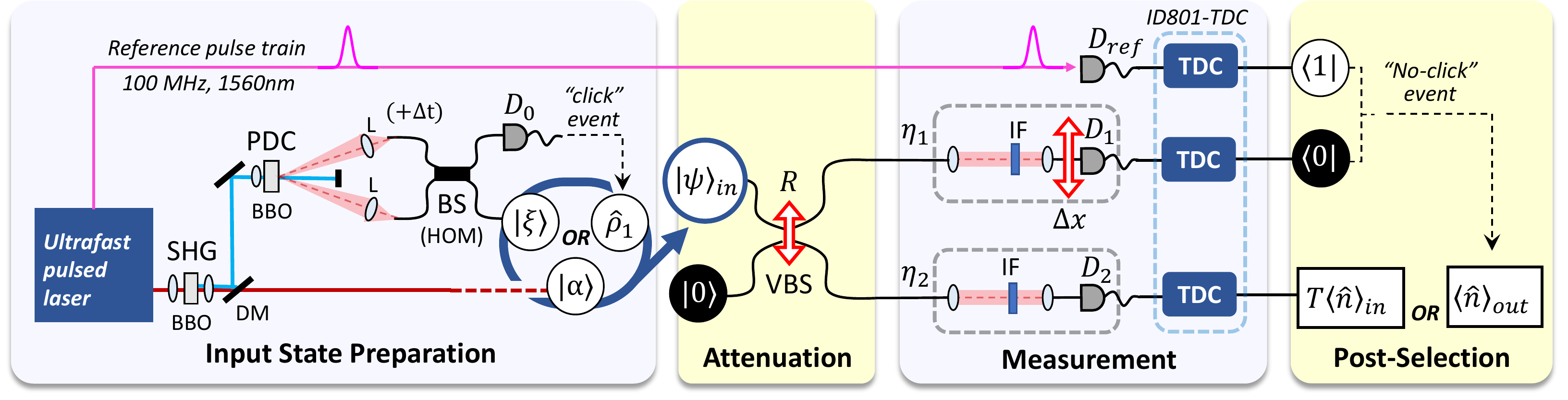}
	\caption{Zero-photon subtraction (ZPS) experiment, in four stages. (1) \textit{Input State Preparation}-- One of three input states is generated, $\ket{\alpha}$, $\ket{\xi}$ or $\hat{\rho}_1$. Coherent states $\ket{\alpha}$ are produced directly by an ultrafast pulsed laser (100 MHz, 780 nm). These pulses also undergo second harmonic generation (SHG) to serve as 390 nm pump pulses for Type-I parametric down-conversion (PDC) using a $\beta$-barium borate (BBO) crystal. The resulting photon pairs are coupled into a Hong-Ou-Mandel (HOM) interferometer to produce either $\ket{\xi}$ or $\hat{\rho}_1$ as described in the text. (2) \textit{Attenuation}-- The input state enters a fiber-based variable beamsplitter (VBS) with reflectance $R$. (3) \textit{Measurement}-- Each VBS output is measured by single-photon detectors $D_1$ and $D_2$, with overall channel efficiencies $\eta_1$ and $\eta_2$. The heralding detector $D_1$ can be translated across the mode of interest by displacement $\Delta x$. Detection events and the $D_{ref}$ reference signal are recorded by time-to-digital converters (TDCs). (4) \textit{Post-selection}-- Time tags from all detections are used to measure $D_2$ counting rates with and without post-selecting on ``no-click'' events at $D_1$. Abbreviations: DM-- dichroic mirror used to isolate UV pump pulses; L-- various lenses; $\Delta t$-- glass wedge time delay; IF-- narrowband interference filters centered near 780 nm.}
	\label{fig:exp}
\end{figure*}

The full ZPS experiment is shown in Figure~\ref{fig:exp}. As summarized in the first panel, one of each type of input state (sub-, super- and Poissonian) is prepared with a combination of standard techniques in quantum optics.

In the case of Poissonian statistics, coherent states $\ket{\alpha}$ are prepared with a mode-locked fiber laser (Menlo Systems C-Fiber 780), which generates a train of ultrashort pulses with a repetition rate of 100 MHz and a center wavelength of 780 nm. These pulses are coupled into a single-mode fiber and attenuated for use as ZPS input states.

Super-Poissonian SMSV states $\ket{\xi}$ are prepared with parametric down-conversion (PDC) and a Hong-Ou-Mandel (HOM) interferometer~\cite{hom_87}. The 780 nm pulse train is first frequency doubled and used as a pump for Type-I PDC using a $\beta$-barium borate (BBO) crystal. The resulting photon pairs are coupled into single-mode fibers, and then combined in a 50-50 fiber coupler serving as the HOM interferometer. The relative time delay $\Delta t$ between photons is controlled with a pair of translating glass wedges before one of the fibers. When $\Delta t =0$, interference ideally produces two disentangled SMSV states in the HOM outputs~\cite{kim_02}. With our low pump power, the PDC photon pair production rate of $\sim10^{-4}$ per pulse ensures we are in the weak squeezing limit where Eq.~\ref{eqn:kx} holds.

Using the same setup with noncollinear PDC and the HOM interferometer, we can also generate heralded single-photon states with sub-Poissonian statistics. First, a large time delay is introduced in one input of the interferometer, eliminating HOM interference. Next, a single-photon detector $D_0$ without photon-number resolution (non-PNR) is coupled to one output. When $D_0$ detects exactly one photon with a ``click,'' the twin photon is heralded in the other mode (offset by the delay $\pm\Delta t$). Alternatively, a ``click'' could result from two photons hitting $D_0$, heralding zero photons in the output. The ideal result is a mixture $\hat{\rho}_1=(1-\beta)\ket{0}\bra{0}+\beta\ket{1}\bra{1}$, with $\beta=2/3$.

ZPS is performed at a variable beamsplitter (VBS), implemented with a tunable fiber coupler~\cite{digshaw_82}. The input state $\ket{\psi}_{in}$ (i.e., $\ket{\alpha}$, $\ket{\xi}$ or $\hat{\rho}_1$) enters one input of the VBS, and the two outputs are routed to detection channels. Each channel includes a free-space U-bench with 25-nm-bandwidth rectangular bandpass filters centered near 780 nm, then coupled into multimode fibers and directed to single-photon counting modules (SPCMs) $D_1$ and $D_2$ (silicon avalanche photodiodes, Excelitas SPCM-AQ4C). The auxiliary heralding detector $D_0$ has a similar channel not shown in Figure~\ref{fig:exp}, with a more narrow 10-nm-bandwidth filter to increase heralding efficiency~\cite{pitt_05,bovino_03}. To serve as a universal clock for all ``click'' and ``no-click'' events, all detection signals are recorded alongside a 100 MHz mode-locking reference signal from an additional detector $D_{ref}$, using time-to-digital converters (TDCs) with 81 ps timebin resolution (IDQuantique, model ID801). All detection events are stored as time tags and processed using the techniques described in Ref.~\cite{nun_21}.

The counting statistics of ZPS states are observed by post-selecting on ``no-click'' events, in which $D_{ref}$ registers a pulse but the heralding detector $D_1$ measures zero photons. After a 20-second exposure, the mean counting rate at $D_2$ is calculated with and without this post-selection. Then the $D_2$ dark count rate ($\sim$80 Hz, after filtering~\cite{wahl_20}) is subtracted from each of these values, and their ratio is taken to obtain $K$. This is repeated for multiple values of VBS reflectance $R$, revealing the behavior of $K(R)$ for each state.

Each stage of the experiment introduces losses which must be taken into account for our analysis. Returning to Figure~\ref{fig:bsinout}, we can group all losses into three distinct channels: the input mode of the main beamsplitter (VBS); the reflected auxiliary mode, containing heralding detector $D_1$; and the transmitted output mode, containing the photon-counting detector $D_2$. Input losses are primarily due to coupling free-space photon pairs from the PDC source into single-mode fibers, as well as fiber connector losses at the VBS. The fiber-coupling efficiency and connector transmission are denoted $\kappa_{\text{PDC}}$ and $\kappa_f$, respectively. Additional losses \textit{after} the VBS are contained in the effective detector efficiencies $\eta_1$ and $\eta_2$, illustrated in the third panel of Fig.~\ref{fig:exp}.

As defined in Eq.~\ref{eqn:kdeff}, $K(R)$ is unaffected by losses in the output mode with detector $D_2$. This can differ for non-PNR detectors as shown in the Appendix, but these effects are negligible in our experiment. However, losses in the heralding mode introduce unwanted noise that alters our counting statistics~\cite{nun_21}. Similarly, input losses introduce noise that alter the photon statistics of the initial states. Even so, the resulting mixed states can be analyzed with the same measurement of $K$, which only depends on diagonal terms $\rho_{nn}$ in a full description of the state. Consequently, we can modify our equations of $K(R)$ to include all losses (see Appendix for details):
\begin{align}
	K_{exp}^{(\alpha)}(R)&=1 \label{eqn:ka2} \\
	K_{exp}^{(\xi)}(R)&\approx 1-\kappa_{\text{PDC}}\kappa_f R \eta_1 \label{eqn:kx2} \\
	K_{exp}^{(\hat{\rho}_1)}(R)&=\frac{1}{1-\beta \kappa_f R \eta_1} \label{eqn:kh2} 
\end{align}

Results for the coherent state $\ket{\alpha}$ remain unchanged from Eq.~\ref{eqn:ka} to Eq.~\ref{eqn:ka2}. In the case of the SMSV (Eq.~\ref{eqn:kx2}), the probability of multipair emission is negligible, and so the HOM interferometer output is dominated by zero- and two-photon terms. Loss before ($\kappa_{\text{PDC}}$) and after the interferometer ($\kappa_f$) introduce a significant single-photon component, but the altered statistics remain super-Poissonian. For the state $\hat{\rho}_1$, the existing single-photon term is similarly reduced by $\kappa_f$ but remains sub-Poissonian. The initial single-photon probability $\beta$ is also degraded by dark counts at $D_0$ and interferometer losses, lowering it from the ideal value of $2/3$. In all cases, finite heralding efficiency $\eta_1$ has the same effect on the measured value of $K$ as the overall input losses.

To experimentally determine coupling values and detector efficiencies in our system, and to align the apparatus for input state preparation, we first perform a series of standard HOM tests~\cite{hom_87} and channel loss measurements. We bypass $D_0$ and the VBS in Figure~\ref{fig:exp} and perform a coincidence measurement with the $D_1$ and $D_2$ channels connected directly to the HOM interferometer outputs. We observe a HOM dip with 98\% visibility with this arrangement. Approximate Klyshko efficiencies~\cite{klyshko_80}, apart from the interferometer coupling efficiency of $\kappa_{\text{PDC}}\approx0.50$, are found to be $\eta_1\approx0.32$ and $\eta_2\approx0.28$. This is consistent with nominal SPCM detector efficiencies of $\sim50\%$ at 780 nm and U-bench transmission of $\sim65\%$ and $\sim60\%$. The values of $\kappa_f\approx0.86$ and $\beta\approx0.38$ are determined in the analysis of the main experiment.

\section{Results and Discussion}\label{sec:res}

\begin{figure}
	\includegraphics[scale=0.45,trim={190 25 225 50},clip]{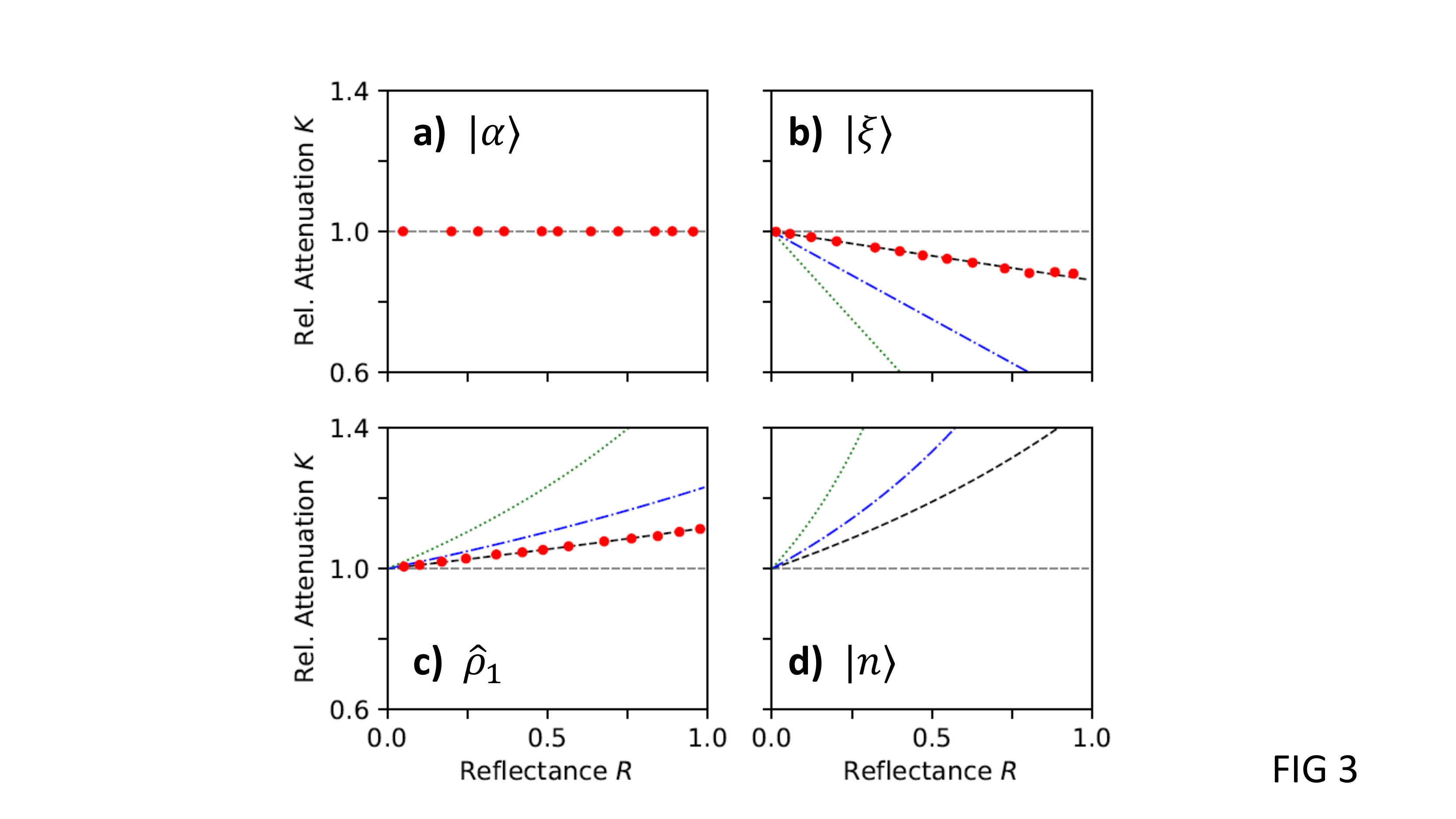}
	\caption{Experimental measurements of relative attenuation $K(R)$ for the three cases of (a) a coherent state $\ket{\alpha}$, (b) a SMSV state $\ket{\xi}$, and (c) a heralded single photon state $\hat{\rho}_1$.  The measured data points (red circles) in plots (a)-(c) show distinct trends that depend on the photon number statistics of the given state: the benchmark case $\ket{\alpha}$ shows $K=1$ for all reflectance $R$, while $\ket{\xi}$ and $\hat{\rho}_1$ have a negative and positive slope, respectively. In panel (b), a fit to the data using Eq.~\ref{eqn:kx2} (black dashed curve) gives the value $(\kappa_{\text{PDC}})(\kappa_f)( \eta_1)= 0.14$, corresponding to an overall loss of 86\%. In panel (c), a similar fit to the data using Eq.~\ref{eqn:kh2} corresponds to an overall loss of 73\% with single-photon probability $\beta\approx0.38$. In both panels (b) and (c), the blue (dot-dashed) and green (dotted) theoretical curves show more pronounced effects of ZPS that would be observed for 50\% and 0\% overall loss, respectively. For comparison, panel (d) shows theoretical attenuation of an ideal Fock state $\ket{n}$ calculated for the same overall loss values as in panel (c).}
	\label{fig:res} 
\end{figure}

Our main results are shown in Figure~\ref{fig:res}. The relative attenuation $K$ induced by ZPS is shown as a function of reflectance $R$ for all three input states. Each state exhibits very distinct behavior in agreement with equations \ref{eqn:ka2}-\ref{eqn:kh2}. For the coherent state shown in Fig.~\ref{fig:res}(a), $K=1$ for all $R$, indicating that ZPS counting rates are identical to those of ordinary attenuation. This case provides a benchmark for our experiment, and can be understood by the well-known fact that a coherent state entering a beamsplitter produces two uncorrelated coherent states in the outputs~\cite{zav_08,gk_04,kim_02}. Consequently, conditional measurements like those in ZPS have no effect on the output state in this case.

In Fig.~\ref{fig:res}(b), measurements of $K$ for the SMSV state $\ket{\xi}$ are shown. The data trend exhibits a negative initial slope in accordance with Eq.~\ref{eqn:qin} $(Q>0)$. This ``photon deficit'' $K<1$  increases linearly with reflectance $R$. This attenuation, however, is limited by heralding efficiency and losses. A fit to the data using Eq.~\ref{eqn:kx2} shows that as $R\rightarrow1$, $K^{(\xi)}_{min}=0.861\pm0.003$. This is consistent with the product of efficiencies $(\kappa_{\text{PDC}})(\kappa_f)( \eta_1)\approx0.14$ (i.e., overall loss of 86\%). For comparison, the two theoretical curves in Fig.~\ref{fig:res}(b) show the stronger attenuation that would be achieved with overall losses of only 50\% and 0\%.

The heralded single photon case $\hat{\rho}_1$ in Fig.~\ref{fig:res}(c) displays essentially opposite behavior. The sub-Poissonian statistics $(Q<0)$ determine a positive initial slope such that $K>1$, and this trend continues for all values of $R$. Note that by our definition of \textit{relative} attenuation, this does not indicate $\langle \hat{n} \rangle_{out}>\langle \hat{n} \rangle_{in}$, and is much different from the ``photon excess'' observed for super-Poissonian states after SPS. The state is still attenuated relative to the input, and this can be seen by comparing the observed $K$ to the theoretically predicted values for ideal Fock states in Fig.~\ref{fig:res}(d), for which $\langle \hat{n} \rangle_{out}=\langle \hat{n} \rangle_{in}$. In contrast to the states $\ket{\alpha}$ and $\ket{\xi}$, however, $K>1$ indicates that the degree of heralded attenuation by ZPS is weaker than that of ordinary attenuation. With previously determined values of $\kappa_f\approx0.86$ and $\eta_1\approx0.32$, which combine to give an overall loss of 73\%, a fit of the data in panel (c) to Eq.~\ref{eqn:kh2} indicates a single photon probability of $\beta\approx0.38$ for our initial state. Two theoretical curves with $\beta=0.38$ and improved overall losses of 50\% and 0\% show more extreme deviations from $K=1$.

Interestingly, the theoretical curves in Figure~\ref{fig:res}(b)-(d) show as detector inefficiency and losses increase, the ZPS statistics for both sub- and super-Poissonian states converge toward the Poissonian case $K=1$. Losses before or after the VBS play an identical role in reshaping $K(R)$ in Eqs.~\ref{eqn:ka2}-\ref{eqn:kh2}, and so we can explain this in two ways. First, the photon number distributions of $\ket{\xi}$ and $\hat{\rho}_1$ tend toward Poissonian statistics after ordinary attenuation, i.e., loss~\cite{hu_07}. As losses before the beamsplitter increase, the observed $K(R)$ values should therefore tend to unity and resemble those of the coherent state. Alternatively, losses before the heralding detector $D_1$ reduce our ability to distinguish ``true'' vacuum in the reflected mode from one or more photons~\cite{nun_21}. As effective efficiency decreases, ``no-click'' events herald a mixture of the desired ZPS state with unwanted noise, becoming identical to ordinary attenuation in the zero-efficiency limit.

\begin{figure}
	\includegraphics[scale=0.45,trim={205 95 205 80},clip]{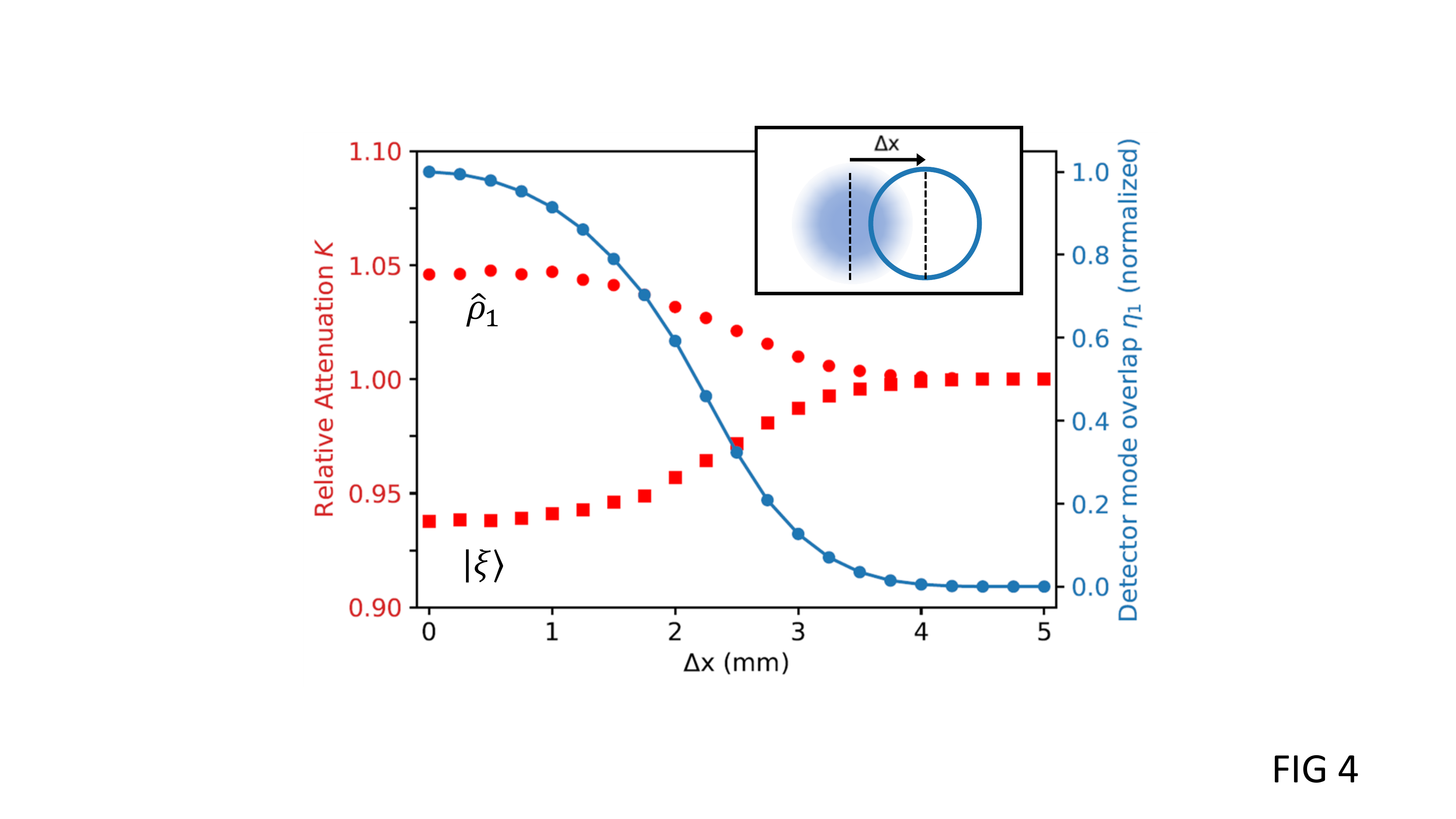}
	\caption{Relative attenuation $K$ for the states $\ket{\xi}$ and $\hat{\rho}_1$ at $R\approx0.5$, measured as the heralding detection channel $D_1$ is moved out of the mode of interest by a distance $\Delta x$, which modulates heralding efficiency $\eta_1$. As $\eta_1$ decreases down to zero (right axis), the relative attenuation values (left axis) converge to $K=1$ as ZPS becomes increasingly ineffective. Here, heralding efficiency is normalized to its maximum value of $\eta_1\approx0.32$.}
	\label{fig:dx}
\end{figure}

Figure~\ref{fig:dx} demonstrates this effect by studying $K(R)$ as detector efficiency $\eta_1$ decreases. Here, the multimode fiber launcher coupled to heralding detector $D_1$ is scanned a distance $\Delta x$ out of the mode of interest within the U-bench (see inset and Fig.~\ref{fig:exp}). The blue curve shows the degree of spatial mode overlap as the fiber is moved, measured separately as the fraction of power coupled from an auxiliary source. As detector mode overlap decreases, it can be seen that the relative attenuations $K^{(\xi)}$ and $K^{(\hat{\rho}_1)}$ (measured at $R\approx0.5$) converge to the Poissonian value of $K=1$. Although we continue to herald on zero photons in $D_1$, these measurements carry less information about the system and ZPS becomes increasingly ineffective.


The results presented in Figs.~\ref{fig:res} and \ref{fig:dx} show a strong connection between the effects of ZPS and the photon number distributions of initial states. Importantly, however, measuring $K(R)$ yields no information about phase or coherence between number states. Consequently, as has been argued for the ``photon excess'' after SPS~\cite{par_07,zav_08}, the effects of ZPS observed here can also be replicated using counting statistics of classical particles and probabilistic subtraction. For ZPS, probability of success drops exponentially with higher numbers of photons, thus shifting the mean of the number distribution downward. Additionally, the ``photon excess'' or ``photon deficit'' exhibited by thermal light undergoing SPS or ZPS can also be understood as intensity fluctuations of the classical electromagnetic field (i.e., correlated intensities at detectors $D_1$ and $D_2$).

In this sense, the ``photon deficit'' observed in ZPS, much like the ``photon excess'' of SPS, is not a purely quantum mechanical effect. Nonetheless, it is important to note that the degree of attenuation measured after ZPS can reveal existing nonclassicality of the input states. For example, our observations confirmed the nonclassical~\cite{gk_04} sub-Poissonian statistics of the heralded single photon $\hat{\rho}_1$. Furthermore, quantum state tomography or some other phase-dependent measurement would reveal that ZPS not only attenuates quantum states, but does so noiselessly (i.e., preserves their coherence)~\cite{saur_21}. This property is exactly what makes ZPS promising for applications in quantum communication~\cite{mic_12,ricky_17,guo_20}.

\section{Conclusions}\label{sec:con}

In summary, we have experimentally demonstrated that the zero-photon subtraction (ZPS) process of Figure~\ref{fig:bsinout} can reduce the mean photon number of quantum optical states, despite no photons being removed from the system. Our experiment tested the effects of ZPS on three unique classes of input states (sub-, super- and Poissonian) using a beamsplitter with variable reflectance $R$. By studying the relative attenuation ratio $K$ as a function of $R$, the observed trends reveal a connection to Mandel's $Q$-parameter in the regime of $R\ll 1$. More precisely, the initial slope of $K(R)$ near $R=0$ is equal to $-Q_{in}$, resulting in distinct behavior for each input state. Consequently, (sub-) super-Poissonian states will be attenuated (less) more by ZPS than by ordinary attenuation with a weakly reflecting beamsplitter. These ZPS effects are complementary to the effects of typical single-photon subtraction (SPS) in the same regime $R\ll 1$. Most notably, super-Poissonian states that exhibit a ``photon excess'' after SPS will also exhibit a unique ``photon deficit'' after ZPS.

These observations were made possible by actively heralding on the detection of zero photons with a single-photon detector~\cite{nun_21}. We further confirmed the need for high efficiency in the heralding mode by measuring the convergence of non-Poissonian attenuation $K$ to the benchmark Poissonian case as losses increased.

Although not revealed by the photon counting measurements reported here, ZPS can preserve the coherence of quantum states~\cite{saur_21}, making it useful for quantum communications as a noiseless attenuator~\cite{mic_12,ricky_17,guo_20}. Our results provide further insight into the nature of this transformation and its relationship to other techniques in quantum state engineering by conditional measurements.

\begin{acknowledgments}
We would like to thank S. U. Shringarpure for many valuable discussions on this topic. This work was supported by the National Science Foundation under Grant No. PHY-2013464.
\end{acknowledgments}

\appendix*

\section{}
This Appendix provides further calculations regarding ZPS with realistic single-photon detectors.

To account for imperfect heralding on zero with finite efficiency $\eta_1$, we can derive an alternate version of Eq.~\ref{eqn:nout}:
\begin{align}\label{eqn:neff}
	\langle \hat{n} \rangle_{out} &= \frac{\text{tr}\left\{\hat{B}\hat\rho\hat{B}^\dagger\hat\Pi^{(NC)}_1\hat{n}_2\right\}}{\text{tr}\left\{\hat{B}\hat\rho\hat{B}^\dagger\hat\Pi^{(NC)}_1\right\}} \\ &= \frac{1-R}{1-R\eta_1}\frac{\sum_n n \rho_{nn}(1-R\eta_1)^n}{\sum_n \rho_{nn}(1-R\eta_1)^n} \nonumber
\end{align}
where the subscripts $1$ and $2$ indicate detection channels $D_1$ and $D_2$, $\hat{B}$ is the unitary beam-splitter operator~\cite{gk_04}, and we have used the standard POVMs for non-PNR detectors for ``click'' $(C)$ and ``no-click'' $(NC)$ events~\cite{stev_13}:
\begin{align}
	\hat\Pi^{(C)}_i &= \mathbbm{1} - \Pi^{(NC)}_i \nonumber \\
	\hat\Pi^{(NC)}_i &= \sum_n (1-\eta_i)^n\ket{n}\bra{n}
\end{align}
for $i=1,2$. Substitution into the definition of $K$ (Eq.~\ref{eqn:kdeff})  yields the same expression, but with the replacement $R\rightarrow R\eta_1$.

To account for losses before the attenuator, we introduce a preceding beamsplitter with transmittance $\kappa$, where the reflected mode is lost to the environment. Tracing over the outputs of both beamsplitters as in~\ref{eqn:neff}, we once again find an expression where losses are included with a simple replacement:
\begin{align}
	K_{exp}(R)= K_{ideal}(\kappa R \eta_1)\label{eqn:keff}
\end{align}
The above relationship gives us Eqs.~\ref{eqn:ka2}-\ref{eqn:kh2}. Mathematically, this means the experimental $K$ and the ideal lossless case have the same general behavior as a function of $R$, except the effective domain is limited from $R\in [0,1)$ to $R_{exp}\in [0,\kappa\eta_1)$.

Finally, we can also account for the lack of PNR capability in the output detector by replacing $\hat{n}_2$ in Eq.~\ref{eqn:neff} with $\hat\Pi^{(C)}_{2}$. The effective relative attenuation becomes:
\begin{equation}
	K_{\text{click}}(R)=\frac{\sum_n\rho_{nn}\left[(1-\kappa R\eta_1)^n-(1-\kappa R\eta_1-T\eta_2)^n\right]}{\left[\sum_n\rho_{nn}(1-\kappa R\eta_1)^n\right]\left[1-\sum_n\rho_{nn}(1-T\eta_2)^n\right]}
\end{equation}
By taking the low-efficiency limit $\eta_2\rightarrow0$ and applying L'Hopital's rule once, we find that $K_{\text{click}}(R)$ converges to $K(R)$. For low mean photon numbers, this converges quickly enough to make the approximation $K_{\text{click}}\approx K$, and the difference is negligible for our experimental conditions.

\end{document}